\documentclass{kluwer}    
\usepackage{epsf}

\begin{document}                                                                                   
\begin{article}
\begin{opening}         
\title{Galaxy Formation: Warm Dark Matter, Missing Satellites, and
the Angular Momentum Problem} 
\author{Martin \surname{G\"{o}tz}\email{gotz@tac.dk}}  
\author{Jesper \surname{Sommer-Larsen}}
\runningauthor{M. G\"{o}tz and J. Sommer-Larsen}
\runningtitle{Warm Dark Matter}
\institute{Theoretical Astrophysics Center, Copenhagen, Denmark}

\begin{abstract}
We present warm dark matter (WDM) as a possible solution to the
missing satellites and angular momentum problem in galaxy formation and
introduce improved initial conditions for numerical simulations of
WDM models, which avoid the formation of unphysical haloes found in
earlier simulations. There is a hint, that the mass function of
satellite haloes has been overestimated so far, pointing to higher
values for the WDM particle mass.
\end{abstract}

\end{opening}           

The cold dark matter (CDM) scenario for structure formation
is well established, and is
particularly successful in explaining the formation of large-scale
structure and galaxies. But there exist problems on small (i.e.\ galactic)
scales, among them: (1) The missing satellites problem: CDM produces
too many small galaxies, e.g.\ about five times as many satellites
as are observed in the local group (\opencite{klypin}; \opencite{moore};
\opencite{kamion}). (2) The angular momentum problem: Galaxies in CDM
simulations consistently have smaller specific angular momenta, and
therefore have smaller disks than what is observed (\opencite{wdm} and
references therein).

One possible solution to these problems is to go from CDM to warm dark matter
(WDM). The free-streaming motion of the WDM particle
reduces power on small scales, but keeps it
unchanged at long wavelengths, not disturbing the predictions of
CDM for the formation of large-scale structure. This leads
to the formation of fewer low-mass systems, explaining the missing
satellites problem, and to fewer merging events, during which subclumps,
which are assembled into galactic disks, would loose orbital angular momentum
and energy by dynamical friction. Typically, WDM particle masses of the
order $1 h^{5/4} \mathrm{ keV}$ (corresponding to a free-streaming mass of
$M_\mathrm{f}\approx 3\cdot 10^{11} h^{-1}M_\odot$ for $\Omega_0 = 0.3$) are
necessary to explain the missing satellites
(\opencite{kamion}; \opencite{colin}; \opencite{wcroft}; \opencite{bode};
\opencite{gover})
and to reduce the discrepancy of the specific angular momenta from a
factor of $10$ or more to about a factor of $2$ (\opencite{wdm}).

Thus to check WDM against observations, it is necessary to study,
how many small haloes there are, and how the large disk galaxies are formed.
This requires numerical simulations, even for
the relatively simple task of calculating the halo mass function in a WDM
scenario. There, structure formation does not follow the
hierarchical picture well-known from CDM, since haloes smaller than
the free-streaming scale form later than the larger ones by non-linear
transfer of power from large to small scales. That means that analytical
schemes (e.g.\ Press-Schechter theory and its variations) can not be
applied to WDM. Figure~1 in \inlinecite{goetz} clearly shows the discrepancy
between Press-Schechter theory and the numerical mass function.

\begin{figure}
\centerline{\epsfxsize=0.86\textwidth \epsfbox{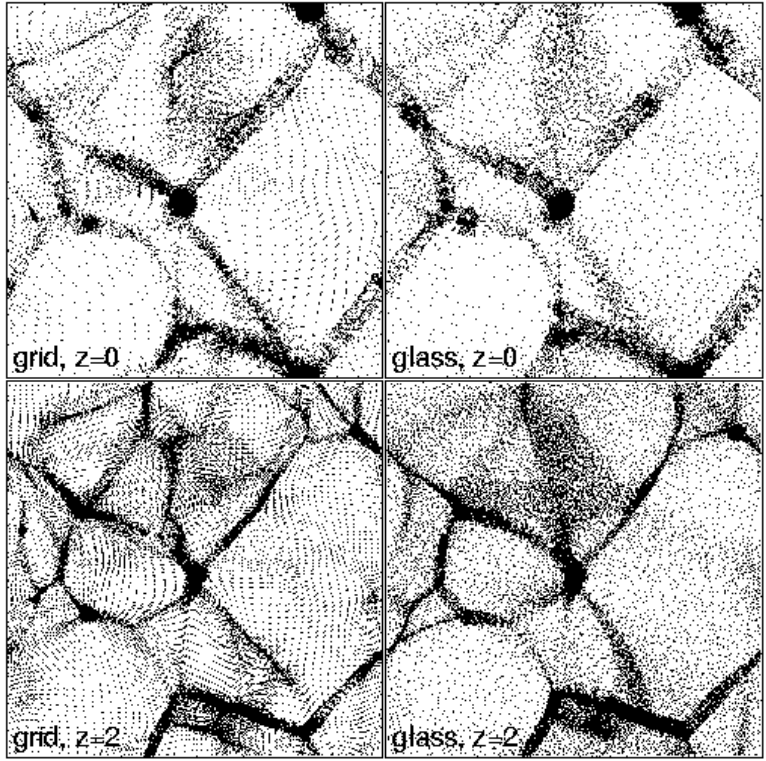}}
\caption[]{Dark matter particles at redshifts $z=0$ and $z=2$ within a slice
of size $10 \times 10 \times 0.325 \left( h^{-1}\mathrm{Mpc}\right)^3$
for two $\Lambda$WDM simulations starting with identical phases from
grid and glass initial conditions, respectively. The regularly spaced,
unphysical haloes
along filaments in grid simulations are clearly visible at $z=0$ (in
particular along the filament running from the halo in the center to the
one close to the lower right corner), and how they form as artifacts due
to trains of particles falling onto the forming filaments at high
redshifts. They are not present in the run starting from a glass.}
\label{glass}
\end{figure}
But numerical simulations with WDM have problems of their own. To check
the effect of reduced power on small scales, they have to be set up
such, that the mean particle separation is (much) smaller than the wavelength,
below which the power spectrum is cut off. Thus initial
displacements and velocities of neighboring particles are highly correlated.
Since at high redshifts, matter moves perpendicularly onto the
forming filaments, there will be ``trains'' of particles
falling onto them
in the usual set-up, where one starts from a regular cubic mesh. They create
unphysical haloes along the filaments, with a separation given by
the grid spacing projected onto the filament \cite{goetz}. These ``beads on
a string'', and how they form, can clearly be seen in the left hand panels of
Figure~\ref{glass}. To avoid this problem, the grid structure needs to
be broken by starting from glass-like initial conditions \cite{glass},
where the particles are irregularly distributed, but still (almost) evenly
spaced.
The right hand panels in Figure~\ref{glass} show the now much smoother
distribution of matter along the filaments.

\begin{figure}
\centerline{\epsfxsize=0.8\textwidth \epsfbox{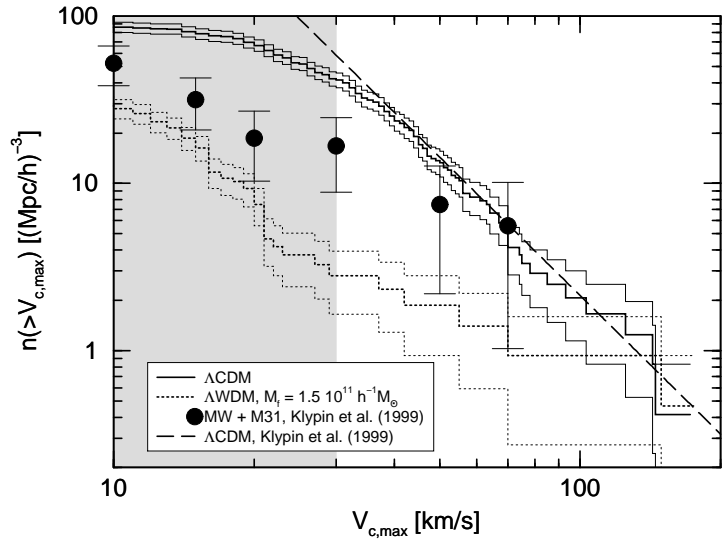}}
\caption[]{The cumulative number density of subhaloes as a function of their
maximum circular velocity $V_{\mathrm{c,max}}$. The subhaloes have to be
located within $400 h^{-1}\mathrm{kpc}$ of the density peak of the host
halo, which are selected to be Milky Way-like with $V_{\mathrm{c,max}}$
between $220$ and $280 \mathrm{\ km}/\mathrm{s}$. The solid and dotted
lines show the numerical results for the $\Lambda$CDM and $\Lambda$WDM
(with a free streaming mass of $M_\mathrm{f}=1.5\cdot 10^{11} h^{-1} M_\odot$)
simulations with glass initial conditions presented in this paper, together
with their $1\sigma$ Poisson errors. The gray area indicates, where these
simulations are incomplete. The observed number density for the Milky Way
and M31 from \inlinecite{klypin} is shown by filled circles, and their
numerical result for a $\Lambda$CDM-model by the dashed line, which agrees
with our result.}
\label{satellites}
\end{figure}
To calculate a more reliable satellite mass function, not affected by
the presence of unphysical haloes, we ran a WDM simulation starting with
glass initial conditions. We chose a $\Lambda$-cosmology ($\Omega_0 = 0.3$,
$\Omega_\Lambda=0.7$, $h=0.65$) with a cluster-abundance normalized power
spectrum ($\sigma_8=1.0$) and a free streaming mass of
$M_\mathrm{f}=1.5\cdot 10^{11} h^{-1} M_\odot$, corresponding to a WDM
particle mass of $1.2 h^{5/4} \mathrm{ keV}$ close to the value found by
other authors. We used the publicly available Hydra code \cite{Hydra}
with $128^3$ dark matter particles in a $10 h^{-1}\mathrm{Mpc}$ box.
A corresponding CDM simulation was also run. Candidate host haloes were
identified with friends-of-friends, and Milky Way-like hosts with a maximum
circular velocity $V_{\mathrm{c,max}}$ between $220$ and
$280 \mathrm{\ km}/\mathrm{s}$ were selected --- $8$ in the $\Lambda$WDM
and $9$ in the $\Lambda$CDM run. Within these haloes, subhaloes were found
with a variation of the HOP algorithm \cite{hop}, and Figure~\ref{satellites}
shows the cumulative number densities of those satellites, which lie
within $400 h^{-1}\mathrm{kpc}$ of the density peak of the host halo, as
a function of their $V_{\mathrm{c,max}}$. As expected, CDM seems to overpredict
the number of satellites compared to the observed distribution for
the Milky Way and M31 \cite{klypin}, but our improved WDM simulations 
without the unphysical haloes now give
numbers which are too low. This hints a lower $M_\mathrm{f}$ and higher
WDM particle mass than has been suggested so far. But that result has to
be taken with caution, since the identification of subhaloes is non-trivial,
and we have not yet checked the effects of different algorithms on it.

In conclusion, WDM can solve the missing satellites and angular momentum
problems, but improved initial conditions starting with a glass-like
distribution of the particles are necessary for numerical studies to avoid
the appearance of unphysical haloes which are purely grid artifacts.
Thus there is the possibility that the satellite mass function has been
overestimated in previous studies, and that the WDM particle mass actually
has to be higher. But it should be mentioned that there are other open
questions for WDM. E.g., high velocity clouds could be the missing
satellites \cite{hvc}, CDM with feedback from star formation may
solve the angular momentum problem (\opencite{cdm}; \opencite{thacker}),
and the late structure formation in WDM can be a problem, if the trend
of finding quasars at ever higher redshifts continues.

\end{article}
\end{document}